\documentclass[12pt]{JHEP3}
\usepackage{mathrsfs}
\usepackage{amsmath,amssymb}
\usepackage{epsfig}
\usepackage{relsize}
\usepackage{graphicx}

\def\tH{\widetilde{{\cal E}}}

\def\sl(2){\alg{sl}(2)}

\def\be{\begin{equation}}
\def\ee{\end{equation}}

\newcommand{\bea}{\begin{eqnarray}}
\newcommand{\eea}{\end{eqnarray}}

\def\a {\alpha}

\def\s {\sigma}

\def\p{\phi}
\def\la{\label}

\def\ov{\over}

\def\bR{{\bf R}}

\def\tp{{\widetilde p}}

\newcommand{\alg}[1]{\mathfrak{#1}}
\newcommand{\su}{\alg{su}}

\newcommand{\psu}{\alg{psu}}

\newcommand{\AdS}{{\rm  AdS}_5\times {\rm S}^5}

\newcommand{\atopfrac}[2]{\genfrac{}{}{0pt}{}{#1}{#2}}

\newcommand{\bem}{\left (\begin{matrix}}
\newcommand{\eem}{\end{matrix} \right )}

\author{Gleb Arutyunov$^a$\footnote{Email: G.Arutyunov@phys.uu.nl, frolovs@maths.tcd.ie} {}\footnote{Correspondent fellow at Steklov
Mathematical Institute, Moscow.}\, and\,  Sergey Frolov$^{b\,
\dagger}$
 \\ $^{a}$ {\it Institute for Theoretical
Physics and Spinoza Institute,\\ ~~Utrecht University, 3508 TD
Utrecht, The Netherlands} \\ $^b$ {\it School of Mathematics,
Trinity College, Dublin 2, Ireland} }

\abstract{ We discuss the states which contribute in the
thermodynamic limit of the mirror theory, the latter is obtained
 from the light-cone gauge-fixed string
theory in the $\AdS$ background by the double-Wick rotation.  We
analyze the Bethe-Yang equations for the mirror theory and
formulate the string hypothesis.  We show that in the
thermodynamic limit solutions of the Bethe-Yang equations arrange
themselves into Bethe string configurations similar to the ones
appearing in the Hubbard model. We also derive a set of equations
describing the bound states  and the Bethe string configurations
of the mirror theory. }

\title{String hypothesis for the $\AdS$ mirror}

\preprint{
          \smaller{\smaller{\smaller{ITP-UU-09-02}}}\\[-.5ex]
          \smaller{\smaller{\smaller{SPIN-09-02}}}\\[-.5ex]
          \smaller{\smaller{\smaller{TCDMATH 09-01}}}}

\begin{document}

\renewcommand{\thefootnote}{\arabic{footnote}}
\setcounter{footnote}{0}


\section{Introduction}
The AdS/CFT correspondence \cite{M} offers new profound insights
into a strong coupling dynamics of gauge theories. In the basic
case of the duality between type IIB superstrings on $\AdS$ and
${\cal N}=4$ SYM one may even hope to find an exact solution of
the tree-level string theory, and, therefore, to solve the dual
gauge theory in the \mbox{`t Hooft} limit. This would be done by
employing the conjectured quantum integrability of the $\AdS$
superstring which is supported by classical integrability
\cite{BPR} of the Green-Schwarz action \cite{MT}, and by one-loop
integrability of the dual gauge theory \cite{MZ,BS1}.

\smallskip

Solving string theory is a multi-step problem. One starts by
imposing the light-cone gauge for the $\AdS$ superstring, and
obtains a 2-d non-linear sigma model defined on a cylinder of
circumference equal to the light-cone momentum $P_+$
\cite{AF0,FPZ}. The gauge-fixed Hamiltonian is equal to $E-J$ and,
therefore, its spectrum determines the spectrum of scaling
dimensions of gauge theory operators. To find the spectrum, one
first takes the decompactification limit \cite{AJK}-\cite{HM},
{\it i.e.} the limit where $P_+$  goes to infinity, while keeping
the string tension $g$ fixed. Then, one is left with a world-sheet
theory on a plane which  has a massive spectrum and well-defined
asymptotic states (particles). This reduces the spectral problem
to finding dispersion relations for particles and the S-matrices
describing their pairwise scattering. Quantum integrability then
implies factorization of multi-particle scattering into a sequence
of two-body events \cite{ZZ}.

\smallskip

To define the S-matrix, one should deal with particles with
arbitrary world-sheet momenta which requires to give up the
level-matching condition. As a result, the manifest $\psu(2|2)
\oplus \psu(2|2) \subset \psu(2,2|4)$ symmetry algebra of the
light-cone string theory gets enhanced by two central charges
\cite{AFPZ}. The same centrally-extended symmetry algebra also
appears in the dual gauge theory  \cite{B}.

\smallskip

An important observation made in \cite{B} is that  the dispersion
relation for fundamental particles  is uniquely  determined by the
symmetry algebra of the model. Moreover, the matrix structure of
their S-matrix is uniquely fixed by the algebra, the Yang-Baxter
equation and the generalized physical unitarity condition \cite{B,
AFZzf, AFtba}.

\smallskip

The S-matrix is thus determined up to an overall  scalar function
$\sigma(p_1,p_2)$ -- the so-called dressing factor \cite{AFS}. Its
functional form was conjectured  in \cite{AFS} by discretizing the
integral equations \cite{KMMZ} describing classical spinning
strings \cite{Gubser:2002tv,FT},  and using insights from gauge
theory \cite{BDS}. It was proposed in \cite{Janik} that the
dressing factor satisfies a crossing equation. Combining the
functional form of the dressing factor together with the first two
known orders in the strong coupling expansion \cite{AFS,HL}, a set
of solutions to the crossing equation in terms of an all-order
strong coupling asymptotic series has been proposed \cite{BHL}.
Opposite to the strong coupling expansion, gauge theory
perturbative expansion of the dressing factor is in powers of $g$
and it has a finite radius of convergence.  An interesting
proposal for the exact dressing factor has been put forward in
\cite{BES}, and passed many tests \cite{RTT}-\cite{BMR}. Thus, one
can adopt the working assumption that the exact dressing factor
and, therefore, the S-matrix are established.

\smallskip

Having found the exact dispersion relation and the S-matrix, the
next step is to determine bound states of the model.  Analysis
reveals that all bound states are those of elementary particles
\cite{D1}, and comprise into the tensor product of two $4Q$-dim
atypical totally symmetric multiplets of the centrally-extended
symmetry algebra $\su(2|2)$ \cite{B2}.

\smallskip

Having understood the spectrum of the light-cone string sigma
model on a plane,   one has to  ``upgrade'' the findings to a
cylinder. All physical string configurations (and dual gauge
theory operators) are characterized by a finite value of $P_+$,
and as such they are excitations of a theory on a cylinder. The
first step in determining the finite-size spectrum is to impose
the periodicity condition on the Bethe wave function. This leads
to a system of equations on the particle momenta known as the
Bethe-Yang equations. In the AdS/CFT context these equations are
usually referred to as the asymptotic Bethe ansatz \cite{S,BS}.
The $\AdS$ string S-matrix has a complicated matrix structure
which results at the end in a set of nested Bethe equations
\cite{B,MM,Le}.

\smallskip

The Bethe-Yang equations determine any power-like $1/P_+$
corrections to energy of multi-particle states. To find  the exact
spectrum for finite values of string tension and $P_+$, one may
try to generalize the thermodynamic Bethe ansatz (TBA), originally
developed for relativistic integrable models \cite{za}, to the
light-cone string theory.

\smallskip

The TBA approach might allow one to relate the exact string
spectrum   to proper  thermodynamic quantities of the mirror
theory obtained from the light-cone string model by means of a
double-Wick rotation. The mirror theory lives on a plane at
temperature $1/P_+$, and, in particular, its Gibbs free energy is
equal to the density of the ground state energy of the string
model. It should be also possible to find the energies of excited
states by analytic continuation of the TBA equations, see e.g
\cite{DT}-\cite{GKV} for some relativistic examples.

\smallskip

Since the light-cone string model is not Lorentz-invariant, the
mirror theory is governed by a different Hamiltonian and therefore
has very different dynamics.  Thus, to implement the TBA approach
one has to study the mirror theory in detail. The first step  in
this direction has been taken in \cite{AFtba}, where the
Bethe-Yang equations for fundamental particles of the mirror model
were derived. Another result of  \cite{AFtba} was the
classification of mirror bound states according to which they
comprise into the tensor product of two $4Q$-dim atypical totally
anti-symmetric multiplets of the centrally-extended algebra
$\su(2|2)$.
 This observation was
used in the derivation \cite{BJ} of the four-loop scaling
dimension of the Konishi operator by means of L\"uscher's formulae
\cite{Lu}.  We consider this derivation as prime evidence for the
validity of the mirror theory approach.

\smallskip

In this paper we take the next step in studying the mirror theory,
and identify the states that contribute in the thermodynamic
limit. We use  the Bethe-Yang equations of \cite{AFtba} and the
fusion procedure, see e.g. \cite{Faddeev}, to write down the
equations for the complete spectrum of the mirror theory. We use
the observation of \cite{B2} that the equations  for auxiliary
roots  can be interpreted as the Lieb-Wu equations for an
inhomogeneous Hubbard model \cite{Lieb}, and notice that the
inhomogeneous Hubbard model becomes homogeneous in the limit of
the infinite real momenta of the mirror particles.  This
observation allows us to formulate the string hypothesis for the
mirror theory. We show that the solutions of the Bethe-Yang
equations in the thermodynamic limit arrange themselves into Bethe
string configurations similar to the ones appearing in the Hubbard
model \cite{Korepin}.  We then derive a set of equations
describing the bound states of the mirror theory and the Bethe
string configurations. These equations can be readily used to
derive a set of TBA equations for the free energy of the mirror
model following a textbook route, see e.g. \cite{Korepin}. The
resulting equations are however complicated and we postpone their
discussion for future publication.

\newpage

\subsection{Bethe-Yang equations}

The  Bethe-Yang equations for fundamental particles and bound states of the mirror theory defined on a circle of large circumference $R$ are derived by using the $\su(2|2)\oplus\su(2|2)$-invariant S-matrix  \cite{AFtba} and the fusion procedure, and are of the form
\begin{eqnarray}\la{BEn2}
1&=&e^{i\tp_{k} R} \prod_ {\textstyle\atopfrac{l=1}{l\neq k}}
^{K^{\mathrm{I}}}S_{\sl(2)}^{Q_kQ_l}(x_{k},x_{l})
\prod_{\a=1}^{2}\prod_{l=1}^{K^{\mathrm{II}}_{(\a)}}\frac{{x_{k}^{-}-y_{l}^{(\a)}}}{x_{k}^{+}-y_{l}^{(\a)}}
\sqrt{\frac{x_k^+}{x_k^-}} \nonumber \\
-1&=&\prod_{l=1}^{K^{\mathrm{I}}}\frac{y_{k}^{(\a)}-x^{-}_{l}}{y_{k}^{(\a)}-x^{+}_{l}}\sqrt{\frac{x_l^+}{x_l^-}}
\prod_{l=1}^{K^{\mathrm{III}}_{(\a)}}\frac{v_{k}^{(\a)}-w_{l}^{(\a)}-\frac{i}{g}}{v_{k}^{(\a)}-w_{l}^{(\a)}+\frac{i}{g}} \\
\nonumber
1&=&\prod_{l=1}^{K^{\mathrm{II}}_{(\a)}}\frac{w_{k}^{(\a)}-v_{l}^{(\a)}+\frac{i}{g}}{w_{k}^{(\a)}-v_{l}^{(\a)}-\frac{i}{g}}
\prod_ {\textstyle\atopfrac{l=1}{l\neq
k}}^{K^{\mathrm{III}}_{(\a)}}\frac{w_{k}^{(\a)}-w_{l}^{(\a)}-\frac{2i}{g}}{w_{k}^{(\a)}-w_{l}^{(\a)}+\frac{2i}{g}}.
\end{eqnarray}
Here $\tp_k$ is the real momentum of a physical mirror particle which can be either a fundamental particle or a $Q$-particle bound state. We will often refer to such a particle as a $Q$-particle, a 1-particle being a fundamental one. Then, $K^{\mathrm{I}}$ is the number of $Q$-particles, and $K^{\mathrm{II}}_{(\a)}$ and $K^{\mathrm{III}}_{(\a)}$ are the numbers of auxiliary roots $y_{k}^{(\a)}$ and $w_{k}^{(\a)}$ of the second and third levels of the nested Bethe ansatz, and $\a=1,2$ because the scattering matrix is the tensor product of the two $\su(2|2)$-invariant S-matrices. We will often refer to $K$'s as to excitation numbers.
 The parameters $v$ are related to $y$ as  $v=y+\frac{1}{y}$. The parameters $x^\pm$ are functions of the string tension $g$, the momentum $\tp$ and the number of constituents $Q$ of a $Q$-particle, and their explicit expressions can be found in appendix \ref{app:disp}, eq.(\ref{xpmtp}).

The function $S_{\sl(2)}^{Q_kQ_l}(x_{k},x_{l})$ is the two-particle scalar S-matrix which describes the scattering of a $Q_k$-particle with momentum $\tp_k$ and a $Q_l$-particle with momentum $\tp_l$ in the $\sl(2)$ sector of the mirror theory. The S-matrix can be found by using the fusion procedure and the following $\sl(2)$ S-matrix of the fundamental particles
\begin{equation}\la{s011}
 S_{\sl(2)}^{11}(x_1,x_2) = \sigma^{-2}_{12} s_{12} \, ,
\quad s_{12} = \frac{x_1^+ - x_2^-}{x_1^-  - x_2^+}\frac{1 - \frac{1}{x_1^- x_2^+}}{1 - \frac{1}{x_1^+ x_2^-}} \, ,
\end{equation}
where $\s_{12}$ is the dressing factor \cite{AFS} that depends on $x^\pm$ and $g$. Its exact form
was conjectured in \cite{BES} but we will not need it here.  For complex values of the momenta $\tp_1$, $\tp_2$ the S-matrix (\ref{s011}) exhibits a pole at $x_1^-=x_2^+$, and
 it is this pole that leads to the existence of a $Q$-particle bound state satisfying the bound state equation  \cite{AFtba}
\bea\la{Qbound}
x_1^- = x_2^+\,,\quad x_2^- = x_3^+\,,\ \ldots, x_{Q-1}^- = x_Q^+\,.
\eea
The equation has a unique solution in the physical region of the mirror theory defined by Im$\, x^\pm <0$ \cite{AFtba}, and it is used in the fusion procedure. It implies that the S-matrix $S_{\sl(2)}^{Q_kQ_l}(x_{k},x_{l})$ depends only on the total real momenta of the $Q$-particles.

Since  $S_{\sl(2)}^{11}(x_{k},x_{l})$ can be also written as
 \bea\la{sp2} S_{\sl(2)}^{11}(x_1,x_2) =
\frac{u_1-u_2 + {2i\ov g}}{u_1-u_2 - {2i\ov g}}\,\times \,  \left({1-{1\ov x_1^+ x_2^-}\ov 1-{1\ov x_1^- x_2^+}}\,\s_{12}\right)^{-2}  \,, \eea
the $Q$-particle bound state equations (\ref{Qbound}) can be cast in the form
\bea\la{Qboundu}
u_j-u_{j+1} - {2i\ov g} =0\ \Longleftrightarrow\ x_j^- = x_{j+1}^+\,, \quad j=1,2,\ldots, Q-1\,.
\eea
Then, the solution to (\ref{Qboundu}) is simply given by the Bethe string
\bea\la{stringQ}
u_j=u+(Q+1-2j)\frac{i}{g}\, ,\quad j=1,\ldots, Q\, ,\quad u\in \bR\, ,
 \eea
where the real rapidity $u$  determines the momentum of the bound state through eq.(\ref{pu}) from appendix \ref{app:disp}.

By taking the complex conjugate of the first Bethe-Yang equation in (\ref{BEn2})  one can easily see that the unitarity of the S-matrix (\ref{s011}) implies that  for
real values of $\tp_k$ the auxiliary roots $y$ either come in pairs $y_2=1/y_1^*$ or lie on unit circle. As a
consequence the variables $v$ and $w$ come in complex conjugate pairs, or are real.

It is the set of Bethe-Yang equations (\ref{BEn2}) we will be using  in the paper to analyze the solutions which contribute in the thermodynamic limit. However, before starting the analysis we would like to show the relation of the last two equations in  (\ref{BEn2}) for the auxiliary roots to the Lieb-Wu equations for the Hubbard model.

\subsection{Relation to the Lieb-Wu equations}

Let us recall that the Lieb-Wu equations are the Bethe equations for the  Hubbard model and have the form \cite{Lieb,Korepin}
\begin{eqnarray}
&&e^{-i\p}e^{i q_k L}=\prod_{l=1}^{M}\frac{\lambda_{l}-\sin q_k -i {U\ov 4}}{\lambda_{l}-\sin q_k +i {U\ov 4}} \label{LW}\,,\\
\nonumber
&&\prod_{l=1}^{N}\frac{\lambda_{k}-\sin q_l -i {U\ov 4}}{\lambda_{k}-\sin q_l +i {U\ov 4}} =
\prod_ {\textstyle\atopfrac{l=1}{l\neq
l}}^{M}\frac{\lambda_{k}-\lambda_l -i {U\ov 2}}{\lambda_{k}-\lambda_l +i {U\ov 2}}\,,
\end{eqnarray}
where $U$ is the coupling constant of the Hubbard model, $q_k$, $k=1,\ldots,N$, and $\lambda_l$, $l=1,\ldots,M$ are charge momenta and spin rapidities, respectively. The arbitrary  constant $\p$ is a twist which has the physical interpretation of the magnetic flux.

 To relate the Bethe-Yang equations (\ref{BEn2}) for  the auxiliary roots to the Lieb-Wu equations\footnote{This relation was first observed in \cite{B2} and is quite natural taking into account that the $\su(2|2)$-invariant S-matrix coincides with Shastry's R-matrix \cite{Shastry}  up to a scalar factor  \cite{B2,MM}. } let us make the following change
 \bea\nonumber y =i e^{-i q}\,,\quad v = 2
\sin q\,,\quad w = 2\lambda\,. \eea Then the second and third equations in
(\ref{BEn2}) can be cast in the form
\bea\nonumber
&&-\prod_{l=1}^{K^{\mathrm{I}}}\frac{e^{-iq_{k}^{(\a)}}+i
x^{-}_{l}}{e^{-iq_{k}^{(\a)}}+ix^{+}_{l}}\sqrt{\frac{x_l^+}{x_l^-}}=
\prod_{l=1}^{K^{\mathrm{III}}_{(\a)}}\frac{\sin
q_{k}^{(\a)}-\lambda_{l}^{(\a)}+\frac{i}{2g}}{\sin
q_{k}^{(\a)}-\lambda_{l}^{(\a)}-\frac{i}{2g}}\,, \\\nonumber
&&
\prod_{l=1}^{K^{\mathrm{II}}_{(\a)}}\frac{\lambda_{k}^{(\a)}-\sin
q_{l}^{(\a)}-\frac{i}{2g}}{\lambda_{k}^{(\a)}-\sin
q_{l}^{(\a)}+\frac{i}{2g}}= \prod_ {\textstyle\atopfrac{l=1}{l\neq
k}}^{K^{\mathrm{III}}_{(\a)}}\frac{\lambda_{k}^{(\a)}-\lambda_{l}^{(\a)}-\frac{i}{g}}{\lambda_{k}^{(\a)}-\lambda_{l}^{(\a)}+\frac{i}{g}}\  .
\eea
Thus, we see that if for each value of $\a$ we identify $g\to 2/U$, $K^{\mathrm{I}}\to L$,
$K^{\mathrm{III}}_{(\a)}\to M$ and $K^{\mathrm{II}}_{(\a)}\to N$, then we get two copies of equations which can be interpreted as the Bethe equations for an inhomogeneous Hubbard model. The inhomogeneities are determined by the real momenta of the physical particles of the mirror theory. One can easily see that in the limit $\tp\to\infty$ the parameters $x^\pm$ behave as $x^+\to 0$, $x^-\to \infty$ and one obtaines the homogeneous Lieb-Wu equations (\ref{LW}) with $\p=(L-2)\pi/2 $.

The relation to the Hubbard model leads us to a natural conjecture that in the thermodynamic limit where $K^{\mathrm{I}}, K^{\mathrm{II}}_{(\a)}, K^{\mathrm{III}}_{(\a)}\to\infty$ the auxiliary roots $y$ and $w$ will arrange themselves in $vw$- and $w$-strings that in the case of the Hubbard model are called the $k$-$\Lambda$ and $\Lambda$ strings \cite{Korepin}.

\section{String hypothesis }

In this section we argue that in the thermodynamic limit
$R, K^{\mathrm{I}}, K^{\mathrm{II}}_{(\a)}, K^{\mathrm{III}}_{(\a)}\to\infty$ with
$K^{\mathrm{I}}/R$ and so on fixed
 the solutions of the Bethe-Yang equations (\ref{BEn2}) are composed of the following four different classes of Bethe strings
\begin{enumerate}

\item A single $Q$-particle with real momentum $\tp_k$ or, equivalently, rapidity $u_k$

\item  A single $y^{(\a)}$-particle corresponding to an auxiliary root $y^{(\a)}$ with $|y^{(\a)}|=1$

\item  $2M$  roots $y^{(\a)}$ and $M$ roots $w^{(\a)}$ combining into a single $M|vw^{(\a)}$-string
\bea\nonumber
&&\hspace{-0.8cm}v_j^{(\a)}=v^{(\a)}+(M+2-2j)\frac{i}{g}\,,\quad v_{-j}^{(\a)}=v^{(\a)}-(M+2-2j)\frac{i}{g}\,,\quad j=1,\ldots,M\,,\\\la{Mvw}
&&\hspace{-0.8cm}w_j^{(\a)}=v^{(\a)}+(M+1-2j)\frac{i}{g}\,,\quad j=1,\ldots,M\,, \quad v\in \bR\,.
\eea
\item  $N$  roots $w^{(\a)}$ combining into a single $N|w^{(\a)}$-string
\bea\la{Mw}
w_j^{(\a)} = w^{(\a)} + {i\ov g}(N+1-2 j )\,,\quad j=1,\ldots ,N\,,\quad w\in {\bf R}\,.~~~~~~
\eea
This  includes $N=1$ which has a single real root $w^{(\a)}$.
\end{enumerate}
According to the string hypothesis for large $R$ almost all
solutions of the Bethe-Yang equations (\ref{BEn2}) are
approximately given by these Bethe strings with corrections
decreasing exponentially in $R$. The last three types are in fact
the same as in the Hubbard model \cite{Korepin}. Every solution of
(\ref{BEn2}) corresponds to a particular  configuration of the
Bethe strings, and consists of
\begin{enumerate}
\item $N_Q$  $\ Q$-particles, $Q=1,2,\ldots,\infty$
\item  $N_y^{(\a)}$  $\ y^{(\a)}$-particles
\item  $N_{M|vw}^{(\a)}$ $\ M|vw^{(\a)}$-strings, $\a=1,2;\ M=1,2,\ldots,\infty$
\item  $N_{N|w}^{(\a)}$ $\ N|w^{(\a)}$-strings, $\a=1,2;\ N=1,2,\ldots,\infty$
\end{enumerate}

\noindent We have infinitely many states of all these kinds in the  thermodynamic limit.
The numbers $N_Q$, $N_y^{(\a)}$,  $N_{M|vw}^{(\a)}$, $N_{N|w}^{(\a)}$ are called the occupation numbers of the root configuration under consideration, and they obey the `sum rules'
\bea\la{sumrules}
K^{{\mathrm{I}}} &=& \sum_{Q=1}^\infty N_Q\,,\\\nonumber
K^{{\mathrm{II}}}_{(\a)} &=& N_y^{(\a)}+\sum_{M=1}^\infty 2M\, N_{M|vw}^{(\a)}\,,\\\nonumber
K^{{\mathrm{III}}}_{(\a)} &=& \sum_{M=1}^\infty M\left( N_{M|vw}^{(\a)} +N_{M|w}^{(\a)}\right)\,.
\eea

Solutions of the Bethe-Yang equations (\ref{BEn2}) with no coinciding roots, and having  excitation numbers  satisfying the following inequalities
\bea\la{ineq}
\sum_{Q=1}^\infty Q\, N_Q\equiv K^{{\mathrm{I}}}_{\rm tot} \ge K^{{\mathrm{II}}}_{(\a)}\ge 2K^{{\mathrm{III}}}_{(\a)}
\eea
are called regular. Solutions which differ by ordering of roots are considered as equivalent.

We expect in analogy with the Hubbard model that each regular solution corresponds to a highest weight state of the four $\su(2)$ subalgebras of the $\su(2|2)\oplus\su(2|2)$ symmetry algebra of the model and vise versa.
The Dynkin labels  are related to the excitation numbers as follows
\bea\nonumber
s_{\a}=K^{{\mathrm{I}}}_{\rm tot}-K^{{\mathrm{II}}}_{(\a)}\,,\quad q_{\a}=K^{{\mathrm{II}}}_{(\a)}-2K^{{\mathrm{III}}}_{(\a)}
\,.
\eea
This follows from the fact that a $Q$-particle is a bound state of $Q$ fundamental particles.

In the remaining part of the section we explain how the Bethe string configurations can be found.

\subsection{$M|vw$-strings}

Let us recall that to find the $Q$-particle bound states one should consider complex values of particle's momenta and take the limit  $R\to\infty$ keeping the numbers $K^{\mathrm{I}}$, $K^{\mathrm{II}}_{(\a)}$ and $K^{\mathrm{III}}_{(\a)}$  of the physical particles and auxiliary roots finite. The Bethe string configurations of the auxiliary roots can be also found in a similar way.

To determine the string configurations of $y_{k}^{(\a)}$ roots we assume that the momenta of physical particles are real, and  take  $K^{\mathrm{I}}$ to infinity keeping $K^{\mathrm{II}}_{(\a)}$ and $K^{\mathrm{III}}_{(\a)}$ finite.

Then, one can easily show that
\bea\nonumber
\prod_{l=1}^{K^{\mathrm{I}}}\frac{y_{k}^{(\a)}-x^{+}_{l}}{y_{k}^{(\a)}-x^{-}_{l}}\sqrt{\frac{x_l^-}{x_l^+}}\ \longrightarrow \ 0 \ \  {\rm if\ \ } |y_{k}^{(\a)}| < 1\,,
\eea
and
\bea\nonumber
\prod_{l=1}^{K^{\mathrm{I}}}\frac{y_{k}^{(\a)}-x^{+}_{l}}{y_{k}^{(\a)}-x^{-}_{l}}\sqrt{\frac{x_l^-}{x_l^+}}\ \longrightarrow \ \infty \ \  {\rm if\ \ } |y_{k}^{(\a)}| > 1\,.
\eea
 If $|y_{k}^{(\a)}|=1$ then the absolute value of the product is equal to 1.

We can consider roots  with $\a=1$, denote them as $y_k$, $v_k$ and $w_k$, and assume without loss of generality that $|y_1|<1$. Then the Bethe-Yang equation for $y_1$ in (\ref{BEn2}) takes the form
\begin{eqnarray}\la{BEy1}
-1=\prod_{l=1}^{K^{\mathrm{I}}}\frac{y_{1}-x^{+}_{l}}{y_{1}-x^{-}_{l}}\sqrt{\frac{x_l^-}{x_l^+}}
\prod_{l=1}^{K^{\mathrm{III}}}\frac{v_{1}-w_{l}+\frac{i}{g}}{v_{1}-w_{l}-\frac{i}{g}}\ \ \longrightarrow\  -1=0\, \times\,
\prod_{l=1}^{K^{\mathrm{III}}}\frac{v_{1}-w_{l}+\frac{i}{g}}{v_{1}-w_{l}-\frac{i}{g}}\,.~~~~~~~
\end{eqnarray}
Thus, to satisfy this equation we must have a root $w_1$ such that
\bea\la{eqv1}
v_{1}-w_{1}-\frac{i}{g} = 0 \ \Longrightarrow\ v_{1}=w_{1}+\frac{i}{g} \,,
\eea
and computing $y_1$ by using $v_1$ we should keep the solution with $|y_1|<1$.

The equation for $w_1$ takes the form
\bea\nonumber
1=\prod_{l=1}^{K^{\mathrm{II}}}\frac{w_{1}-v_{l}-\frac{i}{g}}{w_{1}-v_{l}+\frac{i}{g}}
\prod_ {l=2}^{K^{\mathrm{III}}}\frac{w_{1}-w_{l}+\frac{2i}{g}}{w_{1}-w_{l}-\frac{2i}{g}}\ \longrightarrow\  1={1\ov 0}\times\prod_{l=2}^{K^{\mathrm{II}}}\frac{w_{1}-v_{l}-\frac{i}{g}}{w_{1}-v_{l}+\frac{i}{g}}
\prod_ {l=2}^{K^{\mathrm{III}}}\frac{w_{1}-w_{l}+\frac{2i}{g}}{w_{1}-w_{l}-\frac{2i}{g}}\,.
\eea
We have to assume that there is  a root $v_2$ such that
\bea\la{eqv2}
w_{1}-v_{2}-\frac{i}{g} = 0 \ \Longrightarrow\ v_{2}=w_{1}-\frac{i}{g} \,.
\eea
Otherwise if there is no such $v_2$ then  $w_1-w_2+2i/g = 0$, and, therefore from (\ref{eqv1}), $v_1-w_2+i/g = 0$, and we get into a contradiction with (\ref{BEy1}).

Then the Bethe-Yang equation for $y_2$ in (\ref{BEn2}) acquires  the form
\be\nonumber
-1=\prod_{l=1}^{K^{\mathrm{I}}}\frac{y_{2}-x^{+}_{l}}{y_{2}-x^{-}_{l}}\sqrt{\frac{x_l^-}{x_l^+}}
\prod_{l=1}^{K^{\mathrm{III}}}\frac{v_{2}-w_{l}+\frac{i}{g}}{v_{2}-w_{l}-\frac{i}{g}}\ \ \longrightarrow\  -1=0\times\prod_{l=1}^{K^{\mathrm{I}}}\frac{y_{2}-x^{+}_{l}}{y_{2}-x^{-}_{l}}\sqrt{\frac{x_l^-}{x_l^+}}
\prod_{l=2}^{K^{\mathrm{III}}}\frac{v_{2}-w_{l}+\frac{i}{g}}{v_{2}-w_{l}-\frac{i}{g}}\,.~~~~~~~
\ee
Now, if we take $y_2$ with $|y_2|>1$, then we can satisfy this equation and obtain a $1|vw$-string
\bea\nonumber
v_{1}=v+\frac{i}{g} \,,\ |y_1|<1\,,\quad v_{2}=v-\frac{i}{g}\,,\ |y_2|>1\,, \quad w_1=v\,, \quad v\in \bR\,,~~~~~
\eea
where the roots $y_i$ satisfy $y_2=1/y_1^*$.

On the other hand, if we take $y_2$ with $|y_2|\le 1$, then we get the same conditions we had for $y_1$ in (\ref{BEy1}), and, therefore, there should exist a root $w_2$ such that
\bea\nonumber
v_{2}-w_{2}-\frac{i}{g} = 0 \ \Longrightarrow\ w_{2}=v_{2}-\frac{i}{g}=w_{1}-\frac{2i}{g} \,.
\eea
If we stop here we get a $2|vw$-string with
\bea\nonumber
&&w_1=v+\frac{i}{g}\,,\quad w_{2}=v-\frac{i}{g}\,, \quad v\in \bR\,,\\\nonumber
&&v_{1}=v+\frac{2i}{g} \,,\ |y_1|<1\,,\quad v_{-1}=v-\frac{2i}{g} \,,\ y_{-1}= {1\ov y_1^*}\,, \\\nonumber
&&v_{2}=v\,,\ |y_2|\le1 \,,\quad v_{-2}=v\,,\ y_{-2}={1\ov y_2} \,,
\eea
where we denoted $y_4\equiv y_{-1}$ and $y_3\equiv y_{-2}$.

If we continue the process we get a general $M|vw$-string characterized by the following set of equations
\bea\la{eqMvw}
&&w_j=v+(M+1-2j)\frac{i}{g}\,,\quad j=1,\ldots,M\,, \quad v\in \bR\,,\\\nonumber
&&v_j=v+(M+2-2j)\frac{i}{g}\,,\quad v_{-j}=v-(M+2-2j)\frac{i}{g}\,,\quad j=1,\ldots,M \,,~~~~~~~
\eea
where the corresponding roots $y_j$  and $y_{-j}$ are related as $y_{-j}y_j^*=1$ if $j\ne {M+2\ov2}$, and $y_{{M+2\ov2}}\,y_{-{M+2\ov2}}=1$ (that may happen only for even $M$). Computing them by using $v_j$ and $v_{-j}$  we should keep the solutions with $|y_j|\le1$ and $|y_{-j}|\ge 1$ for $1\le j\le M$.
It is worth mentioning that even though $v_j=v_{-M-2+j}$ for $j=2,3,\ldots,M$, this requirement  guarantees  that all the roots $y_j$ in the string are different. In particular,  for $j=2,3,\ldots,M$ the roots $y_j$ and $y_{-M-2+j}$ are related to each other as $y_jy_{-M-2+j}=1$.
It is also interesting that  Im$(y_1)<0$ for any string. This is the condition the parameters $x^\pm$ have to satisfy because it defines the physical region of the mirror theory.
In general, however, the $y$-roots can take arbitrary values.

\subsection{$M|w$-strings}

As we discussed in the previous subsection if we have a root $y$ with $|y|<1$ (or $|y|>1$) then in the  thermodynamic limit we unavoidably get a $M|vw$-string. So, we just need to consider the case where $|y_1|=1$ that is $v_1$ is real, and takes the values $-2<v_1<2$. Then, taking the limit $K^{\mathrm{II}}_{(\a)}\to\infty$ and keeping $K^{\mathrm{III}}_{(\a)}$ finite, one can easily see that
$$
\prod_{l=1}^{K^{\mathrm{II}}_{(\a)}}\frac{w_{k}^{(\a)}-v_{l}^{(\a)}-\frac{i}{g}}{w_{k}^{(\a)}-v_{l}^{(\a)}+\frac{i}{g}}\ \longrightarrow \ 0 \ \  {\rm if\ \ } {\rm Im}(w_{k}^{(\a)})>0\,,
$$
and
$$
\prod_{l=1}^{K^{\mathrm{II}}_{(\a)}}\frac{w_{k}^{(\a)}-v_{l}^{(\a)}-\frac{i}{g}}{w_{k}^{(\a)}-v_{l}^{(\a)}+\frac{i}{g}}\ \longrightarrow \ \infty \ \  {\rm if\ \ } {\rm Im}(w_{k}^{(\a)})<0\,,
$$
Thus, assuming for definiteness that
Im$(w_1)>0$, we get that the first factor in the third equation in (\ref{BEn2}) is exponentially decreasing, and therefore we should have
$$
w_2 = w_1-{2i\ov g}\,.
$$
Then there are two cases. First we could have
$$
{\rm Im}(w_2)<0\,,
$$
and one can easily check that the equation for $w_2$ is satisfied. The reality condition would also give $w_2 = w_1^*$, and one gets a $2|w$-string
$$
w_1 = w + {i\ov g}\,,\quad w_2 = w - {i\ov g}\,,\quad w\in {\bf R}\,.
$$

If Im$(w_2)>0$ then there is $w_3=w_2 - {2i\ov g}$, and the procedure repeats itself. As a result we get a $M|w$-string
$$
w_j = w + {i\ov g}(M-2 j +1)\,,\quad j=1,\ldots ,M\,,\quad w\in {\bf R}\,,
$$
which is the usual Bethe string.

\section{Bethe-Yang equations for  string configurations}

Next we  express  the Bethe-Yang equations  (\ref{BEn2}) in terms
of real physical momenta  $\tp$ of $Q$-particles, auxiliary
momenta $q$ of $y$-particles with $y=i e^{-iq}$,  real coordinates
$v$ of centers of $vw$-strings,  and  real coordinates $w$ of
centers of $w$-strings.

\subsection{Bethe-Yang equations for $Q$-particles}
The first step is to rewrite the first equation in  (\ref{BEn2}) in terms of
 momenta $q^{(\a)}_k$, $k=1,\ldots,N_y^{(\a)}$ of $y^{(\a)}$-particles, and
coordinates $v^{(\a)}_{k,M}$,  $k=1,\ldots,N_{M|vw}^{(\a)}$ of $vw$-strings.
A simple computation gives
\bea\la{BEQ}
1&=&e^{i\tp_{k} R} \prod_ {\textstyle\atopfrac{l=1}{l\neq k}}
^{K^{\mathrm{I}}}S_{\sl(2)}^{Q_kQ_l}(x_{k},x_{l})
\prod_{\a=1}^{2}\prod_{l=1}^{N_y^{(\a)}}\frac{{x_{k}^{-}-y_{l}^{(\a)}}}{x_{k}^{+}-y_{l}^{(\a)}}
\sqrt{\frac{x_k^+}{x_k^-}}\, \prod_{M=1}^{\infty} \prod_{l=1}^{N_{M|vw}^{(\a)}}S_{xv}^{Q_kM}(x_{k},v_{l,M}^{(\a)})\,,~~~~
\eea
Here the auxiliary S-matrix is given by
\bea\la{Sxv}
S_{xv}^{Q_kM}(x_{k},v_{l,M}^{(\a)}) = \frac{x_{k}^{-}-y_{l,M}^{(\a)+}}{x_{k}^{+}-y_{l,M}^{(\a)+}}\,\frac{x_{k}^{-}-y_{l,M}^{(\a)-}}{x_{k}^{+}-y_{l,M}^{(\a)-}}\,
\frac{x_k^+}{x_k^-}\,\prod_{j=1}^{M-1} \,{u_k^- - v_{l,M}^{(\a)-}-{2i\ov g}j\ov u_k^+ - v_{l,M}^{(\a)+}+{2i\ov g}j}\,,~~~~
\eea
where
\bea
\nonumber
y_{l,M}^{(\a)\pm} =x(v_{l,M}^{(\a)\pm})\,,\quad  v_{l,M}^{(\a)\pm}=v_{l,M}^{(\a)}\pm {i\ov g}M\,,
\eea
and $x(u)$ is defined in (\ref{xu}).

For what follows it is convenient to adopt the following notation
\bea\nonumber
N_y&=&N_y^{(1)} + N_y^{(2)}\,,\quad
y_l =y_{l}^{(1)}\,,\ l=1,\ldots,N_y^{(1)}\,,\quad
 y_{N_y^{(1)} + l} = y_{l}^{(2)}\,,\ l=1,\ldots,N_y^{(2)}\,,\\
v_{k,M}^{(1)}&=& v_{k,M}\,,\quad v_{k,M}^{(2)}= v_{k,-M}\,.
\eea
With this notation the Bethe-Yang equations (\ref{BEQ}) for
$Q$-particles take a slightly simpler form
\bea\la{BEqf}
1=e^{i\tp_{k} R}
 \prod_ {\textstyle\atopfrac{l=1}{l\neq k}}^{K^{\mathrm{I}}}
 S_{\sl(2)}^{Q_kQ_l}(x_{k},x_{l})
\prod_{l=1}^{N_y}\frac{{x_{k}^{-}-y_{l}}}{x_{k}^{+}-y_{l}}\sqrt{\frac{x_k^+}{x_k^-}}\,
\prod_ {\textstyle\atopfrac{M=-\infty}{M\neq 0}}^{\infty}
\prod_{l=1}^{N_{M|vw}}S_{xv}^{Q_kM}(x_{k},v_{l,M})\,,~~~~
\eea
where the auxiliary S-matrix is given by (\ref{Sxv}) with
\bea\la{ypmf}
y_{l,M}^{\pm} =x(v_{l,M}^{\pm})\,,\quad  v_{l,M}^{\pm}=v_{l,M}^{}\pm {i\ov g}|M|\,.
\eea

\subsection{Bethe-Yang equations for $y$-particles}
Next we take a $y^{(\a)}$-particle with the root $y^{(\a)}_k=i e^{-iq^{(\a)}_k}$
and rewrite the second equation in  (\ref{BEn2}) in terms of
coordinates $v^{(\a)}_{k,M}$, $k=1,\ldots,N_{M|vw}^{(\a)}$ of $vw$-strings, and coordinates $w^{(\a)}_{k,N}$, $k=1,\ldots,N_{N|w}^{(\a)}$ of $w$-strings.
The result is
\bea\la{BEy}
-1&=&\prod_{l=1}^{K^{\mathrm{I}}}\frac{y_{k}^{(\a)}-x^{-}_{l}}{y_{k}^{(\a)}-x^{+}_{l}}\sqrt{\frac{x_l^+}{x_l^-}}
 \prod_{M=1}^{\infty} \prod_{l=1}^{N_{M|vw}^{(\a)}}\frac{v_{k}^{(\a)}-v_{l,M}^{(\a)+}}{v_{k}^{(\a)}-v_{l,M}^{(\a)-}} \, \prod_{N=1}^{\infty} \prod_{l=1}^{N_{N|w}^{(\a)}}\frac{v_{k}^{(\a)}-w_{l,N}^{(\a)+}}{v_{k}^{(\a)}-w_{l,N}^{(\a)-}} \,,~~~~~~
\eea where \bea\la{ypm1} w_{l,N}^{(\a)\pm}=w_{l,N}^{(\a)}\pm {i\ov
g}N\,. \eea In fact we get the same equation for any root
$y^{(\a)}_k$, no matter if it is a root of a
 $y$-particle or a $vw$-string.

\subsection{Bethe-Yang equations for $w$-strings}
Now we take a $K|w$-string with the coordinates $w^{(\a)}_{k,K}$.
The last equations in  (\ref{BEn2}) can be written in the form
\begin{eqnarray}\la{BEw3}
-1&=&\prod_{l=1}^{N_y^{(\a)}}\frac{w_{k}^{(\a)}-v_{l}^{(\a)}+\frac{i}{g}}{w_{k}^{(\a)}-v_{l}^{(\a)}-\frac{i}{g}}
\prod_ {N=1}^{\infty} \prod_{l=1}^{N_{N|w}^{(\a)}}\frac{w_{k}^{(\a)}-w_{l,N}^{(\a)+} +{i\ov g}}{w_{k}^{(\a)}-w_{l,N}^{(\a)-}+{i\ov g}}\,\frac{w_{k}^{(\a)}-w_{l,N}^{(\a)+} -{i\ov g}}{w_{k}^{(\a)}-w_{l,N}^{(\a)-}-{i\ov g}}\,.~~~~~
\end{eqnarray}
It is interesting that the equation has no dependence on the coordinates of $vw$-strings.
Multiplying $K$ equations in  (\ref{BEw3})  with the roots $w_{k}^{(\a)}$  that form the $K|w$-string, we get
\begin{eqnarray}\la{BEw}
(-1)^K&=&\prod_{l=1}^{N_y^{(\a)}}\frac{w_{k,K}^{(\a)+}-v_{l}^{(\a)}}{w_{k,K}^{(\a)-}-v_{l}^{(\a)}}
\prod_ {N=1}^{\infty} \prod_{l=1}^{N_{N|w}^{(\a)}}S_{vv}^{KN}(w_{k,K}^{(\a)},w_{l,N}^{(\a)}) \,,~~~~~
\end{eqnarray}
where the auxiliary S-matrix is
\bea
\la{svv}
S_{vv}^{KM}(u,u')&=&
 \frac{u-u' -{i\ov g}(K+M)}{u-u' +{i\ov g}(K+M)}  \frac{u-u' -{i\ov g}(M-K)}{u-u' +{i\ov g}(M-K)}\\\nonumber
 &&~~~~~~~~~~~\times\prod_{j=1}^{K-1}\left(\frac{u-u' -{i\ov g}(M-K+2j)}{u-u' +{i\ov g}(M-K+2j)}\right)^2  \,.~~~~~~\eea

\subsection{Bethe-Yang equations for $vw$-strings}
Finally we take a $K|vw$-string with the coordinates $v^{(\a)}_{k,K}$, and multiply $2K$ equations (\ref{BEy}) with the roots $y_{k}^{(\a)}$  that form the $K|vw$-string.
The resulting equation takes the form
\bea\la{BEvw}
1=\prod_{l=1}^{K^{\mathrm{I}}}S_{xv}^{Q_lK}(x_{l},v_{k,K}^{(\a)})
 \prod_{M=1}^{\infty} \prod_{l=1}^{N_{M|vw}^{(\a)}}S_{vv}^{KM}(v_{k,K}^{(\a)},v_{l,M}^{(\a)}) \,  \prod_{N=1}^{\infty}   \prod_{l'=1}^{N_{N|w}^{(\a)}}S_{vv}^{KN}(v_{k,K}^{(\a)},w_{l',N}^{(\a)}) \,,~~~~~~
\eea
where the auxiliary S-matrix is given by (\ref{svv}).

In fact the  coordinates $w^{(\a)}_{l,N}$ of the $w$-strings appearing in (\ref{BEvw}) can be excluded from the equation if one takes into account that the product of $K$
equations in  (\ref{BEw3})  with  roots $w_{k}^{(\a)}$  that form a $K|vw$-string gives the following equation on the coordinates $v^{(\a)}_{k,K}$ of the $K|vw$-string
\begin{eqnarray}\la{BEvw2}
(-1)^K&=&\prod_{l=1}^{N_y^{(\a)}}\frac{v_{k,K}^{(\a)+}-v_{l}^{(\a)}}{v_{k,K}^{(\a)-}-v_{l}^{(\a)}}
\prod_ {N=1}^{\infty} \prod_{l=1}^{N_{N|w}^{(\a)}}S_{vv}^{KN}(v_{k,K}^{(\a)},w_{l,N}^{(\a)}) \,.~~~~~
\end{eqnarray}
Thus, (\ref{BEvw}) and (\ref{BEvw2}) lead to the following equation
\bea\la{BEvw3}
(-1)^K=\prod_{l=1}^{K^{\mathrm{I}}}S_{xv}^{Q_lK}(x_{l},v_{k,K}^{(\a)})\prod_{l=1}^{N_y^{(\a)}}\frac{v_{k,K}^{(\a)-}-v_{l}^{(\a)}}{v_{k,K}^{(\a)+}-v_{l}^{(\a)}}
 \prod_{M=1}^{\infty} \prod_{l=1}^{N_{M|vw}^{(\a)}}S_{vv}^{KM}(v_{k,K}^{(\a)},v_{l,M}^{(\a)})~~~~~~
\eea
which has no dependence on the coordinates of $w$-strings.

The set of the equations (\ref{BEQ}),  (\ref{BEy}),  (\ref{BEvw}), (\ref{BEw}) can be used to derive the TBA equations for the free energy of the mirror model. These equations and their consequences will be discussed in our forthcoming  publication.

\section*{Acknowledgements}
The work of G.~A. was supported in part by the RFBR grant
08-01-00281-a, by the grant NSh-672.2006.1, by NWO grant 047017015
and by the INTAS contract 03-51-6346. The work of S.F. was
supported in part by the Science Foundation Ireland under Grant
No. 07/RFP/PHYF104.

\section{Appendix: Mirror dispersion and parametrizations}\la{app:disp}

The dispersion relation in any quantum field theory can be found
by analyzing the pole structure of the corresponding two-point
correlation function. Since the correlation function can be
computed in Euclidean space, both dispersion relations in the
original theory with $H$ and in the mirror one with $ \widetilde
H$ are obtained from the following expression \bea\la{disrele}
H_{{\rm E}}^2 + 4g^2\sin^2{p_{{\rm E}}\ov 2} +Q^2\,, \eea which
appears in the pole of the 2-point correlation function. Here  we consider the light-cone gauge-fixed string
theory on $\AdS$ which has the Euclidean dispersion relation
(\ref{disrele}) for $Q$-particle bound states in the decompactification limit $L\equiv
P_+\to\infty$. The parameter $g$ is  the
string tension, and is related to the 't Hooft coupling $\lambda$
of the dual gauge theory as $g={\sqrt\lambda\ov 2\pi}$.

\smallskip

Then the dispersion relation in the original theory follows from
the analytic continuation (see also \cite{AJK})
 \bea H_{{\rm E}}\to -i H\,,~~\quad
p_{{\rm E}}\to p\ ~~\Rightarrow~~\
H^2 = Q^2+4g^2\sin^2{p\ov 2} \,, \label{ordis}
\eea
and the mirror one from
\bea \label{mirror}~~~~~~~~H_{{\rm E}}\to \widetilde p\,,~~~\quad
p_{{\rm E}}\to i\widetilde H\ ~~\Rightarrow\ ~~\widetilde H = 2\,
{\rm arcsinh}\Big( {1\ov 2g}\sqrt{Q^2+\widetilde p^2}\Big)\,.
\eea Comparing these formulae, we see that $p$ and $\widetilde p$
are related by the following analytic continuation
\bea\la{ancon}
p \to  2i\, {\rm arcsinh}\Big( {1\ov 2g}\sqrt{Q^2+\widetilde
p^2}\Big)\,,~~\quad~~ H=\sqrt{Q^2+4g^2\sin^2{p\ov 2}}\to
i\widetilde p\,. \eea

In what follows we need to know how the parameters $x^{Q\pm}$
which satisfy the relations
\bea
x^{Q+} + {1\ov x^{Q+}} -x^{Q-} - {1\ov x^{Q-}} = 2i{Q\ov g}\,,\quad {x^{Q+} \ov x^{Q-}} =e^{ip}\,,
\eea
 are expressed through $\tp$. By using
formulae (\ref{ancon}), we find
\bea\la{xpmtp}
x^{Q\pm}(\widetilde p)= {1\ov
2g}\left(\sqrt{1 +{4g^2\ov Q^2+\widetilde p^2}}\ \mp\ 1\right)\left(
\widetilde p -i Q\right)\, ,\eea
where we fix the sign of the square root from the conditions
\bea
{\rm Im}\, \left( x^{Q+} \right) <0\,,\quad {\rm Im}\, \left( x^{Q-} \right) <0\ \ {\rm for}\ \ \tp\in \bR\,.
\eea
As a consequence, one gets
$$
ix^{Q-} -ix^{Q+}= {i\ov g} \left(\widetilde p - iQ\right)\,,\quad  x^{Q+}x^{Q-}=\frac{\widetilde{p}-i Q}{\widetilde{p}+i Q} \,,
$$
which implies that $|x^{Q+}x^{Q-}|=1$ and $|x^{Q+}|<|x^{Q-}|$ for $\widetilde{p}$ real. Also one has \bea
x^{Q\pm}(-\tilde{p})=-\frac{1}{x^{Q\mp}(\tilde{p})}\, ,  \qquad (x^{Q\pm}(\tp))^*=\frac{1}{x^{Q\mp}(\tp^*)}\eea

Note that these relations are well-defined for real $\widetilde{p}$, but one
should use them with caution for complex values of $\widetilde{p}$.  Our choice of the square root cut agrees with the one used in Mathematica: it goes over the negative semi-axes.

It what follows it will be often convenient to use the $u$-rapidity variables defined by
\bea\nonumber
&&u = {1\ov 2}\left(x^{Q+} + {1\ov x^{Q+}} +x^{Q-} + {1\ov x^{Q-}} \right) = x^{Q+} + {1\ov x^{Q+}} -i{Q\ov g}= x^{Q-} + {1\ov x^{Q-}} +i{Q\ov g}\,,\\\la{uxpm}
&&u^{Q+} = x^{Q+} + {1\ov x^{Q+}}= u +i{Q\ov g}\,,\quad u^{Q-} = x^{Q-} + {1\ov x^{Q-}}= u-i{Q\ov g}\,.~~~~
\eea
 The $u$-variable is expressed in terms of $\widetilde{p}$ as
\bea
u(\tp)=\frac{\tp}{g}\sqrt{1+\frac{4g^2}{Q^2+\tp^2}}\,,
\eea
and it is an odd function of $\tp$. The parameters $x^{Q\pm}$ and $\tp$ are expressed in terms of $u$  as follows
\bea
x^{Q+}(u)&=& {1\ov 2}\left( u+ {iQ\ov g} - i \sqrt{4 - \left(u+ {iQ\ov g}\right)^2}\right)\,,\\
x^{Q-}(u)&=& {1\ov 2}\left( u- {iQ\ov g} - i \sqrt{4 - \left(u- {iQ\ov g}\right)^2}\right)\,,\\\la{pu}
\tp^Q(u)&=& {i g\ov 2}\left(  \sqrt{4 - \left(u+ {iQ\ov g}\right)^2}-\sqrt{4 - \left(u- {iQ\ov g}\right)^2}\right)\,.
\eea
Here the cuts in the $u$-plane run from $\pm\infty$ to $\pm2 \pm {i Q\ov g}$ along
the horizontal lines. The $u$-plane with the cuts is mapped onto the region ${\rm Im}\, \left( x^{Q\pm} \right) <0$ which is the physical region of the mirror theory, and therefore it is natural to expect that the $u$-plane should be used in all the considerations. To describe bound states for all values of $\tp$ one should also add either the both lower or both upper edges of the cuts to the $u$-plane. They correspond ${\rm Im}\, \left( x^{Q\pm} \right) =0$. This breaks the parity invariance of the model.

The energy of a $Q$-particle is expressed in terms of $u$ as follows
\bea\la{Hu}
\tH^Q(u)=\log {x^{Q-}\ov x^{Q+}}=2 {\rm arcsinh}\left(\frac{\sqrt{\left(u^2+\sqrt{\frac{\left(u^
   2-4\right)^2 g^4+2 Q^2
   \left(u^2+4\right)
   g^2+Q^4}{g^4}}-4\right) g^2+Q^2}}{2
   \sqrt{2} g}\right)\,,~~~~
\eea
and it is positive for real values of $u$.

It would be also convenient to introduce the function
\bea\la{xu}
x(u)&=& {1\ov 2}\left( u - i \sqrt{4 - u^2}\right)\,,
\eea
with the cuts in the $u$-plane run from $\pm\infty$ to $\pm2$ along
the real  lines, so that
\bea
x^{Q+}(u) = x(u+ {iQ\ov g})\,,\quad x^{Q-}(u) = x(u- {iQ\ov g})\,.
\eea
Also one has \bea\nonumber
x^{Q\pm}(-u)=-\frac{1}{x^{Q\mp}(u)}\, , \ x(-u)=-\frac{1}{x(u)}\, , \quad
(x^{Q\pm}(u))^*=\frac{1}{x^{Q\mp}(u^*)}\, , \
(x(u))^*=\frac{1}{x(u^*)}\,,\eea
and
 \bea
\tp(-u)=-\tp(u)\, , \qquad
(\tp(u))^*=\tp(u^*)\,.\eea



\begin{thebibliography}{20}
\bibitem{M}
  J.~M.~Maldacena,
  ``The large N limit of superconformal field theories and supergravity,''
  Adv.\ Theor.\ Math.\ Phys.\  {\bf 2} (1998) 231
  [Int.\ J.\ Theor.\ Phys.\  {\bf 38} (1999) 1113]
  [arXiv:hep-th/9711200].

\bibitem{BPR}
I.~Bena, J.~Polchinski and R.~Roiban, ``Hidden symmetries of the
$\AdS$ superstring,'' Phys.\ Rev.\ D {\bf 69} (2004) 046002,
hep-th/0305116.

\bibitem{MT}
  R.~R.~Metsaev and A.~A.~Tseytlin,
  ``Type IIB superstring action in AdS(5) x S(5) background,''
  Nucl.\ Phys.\  B {\bf 533} (1998) 109
  [arXiv:hep-th/9805028].

\bibitem{MZ}
  J.~A.~Minahan and K.~Zarembo,
  ``The Bethe-ansatz for N = 4 super Yang-Mills,''
  JHEP {\bf 0303} (2003) 013, hep-th/0212208.

\bibitem{BS1}
  N.~Beisert and M.~Staudacher,
  ``The N=4 SYM Integrable Super Spin Chain,''
  Nucl.\ Phys.\  B {\bf 670} (2003) 439
  [arXiv:hep-th/0307042].

\bibitem{AF0}
  G.~Arutyunov and S.~Frolov,
  ``Integrable Hamiltonian for classical strings on $\AdS$,''
  JHEP {\bf 0502} (2005) 059, hep-th/0411089.

\bibitem{FPZ}
  S.~Frolov, J.~Plefka and M.~Zamaklar,
  ``The $\AdS$ superstring in light-cone gauge and its Bethe
  equations,''
  J.\ Phys.\ A  {\bf 39} (2006) 13037, hep-th/0603008.

\bibitem{AJK}
  J.~Ambjorn, R.~A.~Janik and C.~Kristjansen,
  ``Wrapping interactions and a new source of corrections to the spin-chain /
  string duality,''
  Nucl.\ Phys.\ B {\bf 736} (2006) 288, hep-th/0510171.


\bibitem{Klose}
  T.~Klose and K.~Zarembo,
  ``Bethe ansatz in stringy sigma models,''
  J.\ Stat.\ Mech.\  {\bf 0605} (2006) P006, hep-th/0603039.

\bibitem{Arutyunov:2006iu}
  G.~Arutyunov and S.~Frolov,
  ``On AdS(5) x S**5 string S-matrix,''
  Phys.\ Lett.\  B {\bf 639} (2006) 378
  [arXiv:hep-th/0604043].

\bibitem{HM}
  D.~M.~Hofman and J.~M.~Maldacena,
  ``Giant magnons,''
  J.\ Phys.\ A  {\bf 39}, 13095 (2006), hep-th/0604135.

\bibitem{ZZ}
  A.~B.~Zamolodchikov and A.~B.~Zamolodchikov,
  ``Factorized S-matrices in two dimensions as the exact solutions of  certain
  relativistic quantum field models,''
  Annals Phys.\  {\bf 120} (1979) 253.

  \bibitem{AFPZ}
  G.~Arutyunov, S.~Frolov, J.~Plefka and M.~Zamaklar,
  ``The off-shell symmetry algebra of the light-cone $\AdS$
  superstring,''
  J.\ Phys.\ A  {\bf 40} (2007) 3583, hep-th/0609157.

 \bibitem{B}
  N.~Beisert,
  ``The su(2$|$2) dynamic S-matrix,''
  Adv.\ Theor.\ Math.\ Phys.\  {\bf 12}, 945 (2008), hep-th/0511082.


  \bibitem{AFZzf}
  G.~Arutyunov, S.~Frolov and M.~Zamaklar,
  ``The Zamolodchikov-Faddeev algebra for AdS(5) x S**5 superstring,''
  JHEP {\bf 0704} (2007) 002
  [arXiv:hep-th/0612229].

 \bibitem{AFtba}
  G.~Arutyunov and S.~Frolov,
  ``On String S-matrix, Bound States and TBA,''
  JHEP {\bf 0712} (2007) 024
  [arXiv:0710.1568 [hep-th]].



\bibitem{AFS}
  G.~Arutyunov, S.~Frolov and M.~Staudacher,
   ``Bethe ansatz for quantum strings,''
  JHEP {\bf 0410}, 016 (2004), hep-th/0406256;



  \bibitem{KMMZ}
    V.~A.~Kazakov, A.~Marshakov, J.~A.~Minahan and K.~Zarembo,
  ``Classical / quantum integrability in AdS/CFT,''
  JHEP {\bf 0405} (2004) 024
  [arXiv:hep-th/0402207].


\bibitem{Gubser:2002tv}
  S.~S.~Gubser, I.~R.~Klebanov and A.~M.~Polyakov,
  ``A semi-classical limit of the gauge/string correspondence,''
  Nucl.\ Phys.\  B {\bf 636} (2002) 99, hep-th/0204051.

\bibitem{FT}
S.~Frolov and A.A.~Tseytlin, ``Semiclassical quantization of
rotating superstring in $\AdS$,''
  JHEP {\bf 0206} (2002) 007, hep-th/0204226 \,
$\bullet$ ``Multi-spin string solutions in $\AdS$,''
  Nucl.\ Phys.\ B {\bf 668} (2003) 77, hep-th/0304255 \,

\bibitem{BDS}
  N.~Beisert, V.~Dippel and M.~Staudacher,
  ``A novel long range spin chain and planar N = 4 super Yang-Mills,''
  JHEP {\bf 0407} (2004) 075
  [arXiv:hep-th/0405001].



\bibitem{Janik}
  R.~A.~Janik,
  ``The $\AdS$ superstring worldsheet S-matrix and crossing symmetry,''
  Phys.\ Rev.\ D {\bf 73} (2006) 086006,
  hep-th/0603038.



\bibitem{HL}
  R.~Hernandez and E.~Lopez,
  ``Quantum corrections to the string Bethe ansatz,''
  JHEP {\bf 0607} (2006) 004, hep-th/0603204.

\bibitem{BHL}
  N.~Beisert, R.~Hernandez and E.~Lopez,
  ``A crossing-symmetric phase for AdS(5) x S**5 strings,''
  JHEP {\bf 0611} (2006) 070
  [arXiv:hep-th/0609044].


 \bibitem{BES}
  N.~Beisert, B.~Eden and M.~Staudacher,
  ``Transcendentality and crossing,''
  J.\ Stat.\ Mech.\  {\bf 0701} (2007) P021
  [arXiv:hep-th/0610251].


    \bibitem{RTT}
  R.~Roiban, A.~Tirziu and A.~A.~Tseytlin,
  ``Two-loop world-sheet corrections in $\AdS$ superstring,''
  JHEP {\bf 0707} (2007) 056
  [arXiv:0704.3638 [hep-th]].

  \bibitem{KlMMZ}
  T.~Klose, T.~McLoughlin, J.~A.~Minahan and K.~Zarembo,
  ``World-sheet scattering in AdS(5) x S**5 at two loops,''
  JHEP {\bf 0708} (2007) 051
  [arXiv:0704.3891 [hep-th]].


   \bibitem{Bern}
 Z.~Bern, M.~Czakon, L.~J.~Dixon, D.~A.~Kosower and V.~A.~Smirnov,
  ``The Four-Loop Planar Amplitude and Cusp Anomalous Dimension in Maximally
  Supersymmetric Yang-Mills Theory,''
  Phys.\ Rev.\  D {\bf 75} (2007) 085010
  [arXiv:hep-th/0610248].

  \bibitem{BMR}
  N.~Beisert, T.~McLoughlin and R.~Roiban,
  ``The Four-Loop Dressing Phase of N=4 SYM,''
  Phys.\ Rev.\  D {\bf 76} (2007) 046002
  [arXiv:0705.0321 [hep-th]].

  \bibitem{D1}
  N.~Dorey,
  ``Magnon bound states and the AdS/CFT correspondence,''
  J.\ Phys.\ A  {\bf 39} (2006) 13119
  [arXiv:hep-th/0604175].

 \bibitem{B2}
 N.~Beisert,
  ``The Analytic Bethe Ansatz for a Chain with Centrally Extended su(2|2)
  Symmetry,''
  J.\ Stat.\ Mech.\  {\bf 0701} (2007) P017
  [arXiv:nlin/0610017].


\bibitem{S}
  M.~Staudacher,
  ``The factorized S-matrix of CFT/AdS,''
  JHEP {\bf 0505} (2005) 054, hep-th/0412188;


\bibitem{BS}
N.~Beisert and M.~Staudacher,
  ``Long-range PSU(2,2|4) Bethe ansaetze for gauge theory and strings,''
  Nucl.\ Phys.\  B {\bf 727} (2005) 1
  [arXiv:hep-th/0504190].


\bibitem{MM}
M.~J.~Martins and C.~S.~Melo,
  ``The Bethe ansatz approach for factorizable centrally extended S-matrices,''
  Nucl.\ Phys.\  B {\bf 785} (2007) 246
  [arXiv:hep-th/0703086].


\bibitem{Le}
  M.~de Leeuw,
  ``Coordinate Bethe Ansatz for the String S-Matrix,''
  J.\ Phys.\ A  {\bf 40} (2007) 14413
  [arXiv:0705.2369 [hep-th]].

  \bibitem{za}
  A.~B.~Zamolodchikov,
  ``THERMODYNAMIC BETHE ANSATZ IN RELATIVISTIC MODELS. SCALING THREE STATE POTTS AND LEE-YANG MODELS,''
  Nucl.\ Phys.\  B {\bf 342} (1990) 695.


\bibitem{BJ}
  Z.~Bajnok and R.~A.~Janik,
  ``Four-loop perturbative Konishi from strings and finite size effects for
  multiparticle states,''
  Nucl.\ Phys.\  B {\bf 807} (2009) 625
  [arXiv:0807.0399 [hep-th]].


\bibitem{Lu}
  M.~Luscher,
  ``Volume Dependence Of The Energy Spectrum In Massive Quantum Field Theories.
  1. Stable Particle States,''
  Commun.\ Math.\ Phys.\  {\bf 104} (1986) 177.
  \,
$\bullet$
  ``Volume Dependence of the Energy Spectrum in Massive Quantum Field Theories.
  2. Scattering States,''
  Commun.\ Math.\ Phys.\  {\bf 105} (1986) 153.


  \bibitem{DT}
  P.~Dorey and R.~Tateo,
  ``Excited states by analytic continuation of TBA equations,''
  Nucl.\ Phys.\  B {\bf 482} (1996) 639
  [arXiv:hep-th/9607167].

  \bibitem{BH}
  J.~Balog and A.~Hegedus,
  ``TBA equations for excited states in the O(3) and O(4) nonlinear
  sigma-model,''
  J.\ Phys.\ A  {\bf 37} (2004) 1881
  [arXiv:hep-th/0309009].

  \bibitem{Heg}
  A.~Hegedus,
  ``Nonlinear integral equations for finite volume excited state energies  of
  the O(3) and O(4) nonlinear sigma-models,''
  J.\ Phys.\ A  {\bf 38} (2005) 5345
  [arXiv:hep-th/0412125].

  \bibitem{Teschner:2007ng}
  J.~Teschner,
  ``On the spectrum of the Sinh-Gordon model in finite volume,''
  Nucl.\ Phys.\  B {\bf 799} (2008) 403
  [arXiv:hep-th/0702214].

\bibitem{GKV}
  N.~Gromov, V.~Kazakov and P.~Vieira,
  ``Finite Volume Spectrum of 2D Field Theories from Hirota Dynamics,''
  arXiv:0812.5091 [hep-th].


  \bibitem{Faddeev}
  L.~D.~Faddeev,
  ``How Algebraic Bethe Ansatz works for integrable model,''
  arXiv:hep-th/9605187.

 \bibitem{Lieb}
  E.~H.~Lieb and F.~Y.~Wu,
  ``Absence Of Mott Transition In An Exact Solution Of The Short-Range,
  One-Band Model In One Dimension,''
  Phys.\ Rev.\ Lett.\  {\bf 20} (1968) 1445.

\bibitem{Korepin}
F.H.L. Essler, H. Frahm, F. G\"ohmann, A. Kl\"umper, V.E. Korepin,

``The One-Dimensional Hubbard Model,'' Cambridge University Press (2005).

\bibitem{AFs}
  G.~Arutyunov and S.~Frolov,
  ``The S-matrix of String Bound States,''
  Nucl.\ Phys.\  B {\bf 804} (2008) 90
  [arXiv:0803.4323 [hep-th]].

 \bibitem{Le2}
  M.~de Leeuw,
  ``The Bethe Ansatz for AdS5 x S5 Bound States,''
  arXiv:0809.0783 [hep-th].

 \bibitem{Shastry} B.~S.~Shastry, ``Exact integrability of the
 one-dimensional Hubburd-model", Phys.Rev.Lett {\rm 56} (1986)
 2453.

\end{thebibliography}
\end{document}